**Nematic and Smectic phases with Proper Ferroelectric Order**

*Grant J. Strachan,\* Ewa Górecka, Jadwiga Szydłowska, Anna Makal, Damian Pociecha*

G.J. Strachan, E. Górecka, J. Szydłowska, A. Makal, D. Pociecha

Faculty of Chemistry, University of Warsaw, Zwirki i Wigury 101, 02-089 Warsaw, Poland

E-mail: g.strachan@chem.uw.edu.pl

A material showing a sequence of three ferroelectric liquid crystalline phases below the paraelectric nematic phase has been synthesized and studied. The polar order of molecules appearing due to the dipole-dipole interactions in the $N_F$ phase is preserved also in the smectic phases: orthogonal $SmA_F$ and tilted $SmC_F$. The ferroelectric ground state of both smectic phases is confirmed by their second harmonic generation activity and polarization switching. In the $SmC_F$ phase the polarization becomes oriented to the electric field by decreasing the tilt angle to zero. Although both smectic phases are ferroelectric in nature, their dielectric response is found to be very different.

## 1. Introduction

Ferroelectricity is a material property that refers to the presence of spontaneous electric polarization which is reversible on the application of an electric field. It was first discovered in Rochelle salt by J. Valasek in 1921,[1] and today, there are only around 300 known ferroelectric crystals, making it still a relatively uncommon property for crystals. Ferroelectricity is also relatively rare in soft matter. Ferroelectric polymers maintain a permanent electric polarization due to the all-*trans* conformation of polymer chains and thus parallel ordering of transverse dipole moments, and the most studied example is polyvinylidene fluoride.[2] In the 1970s, ferroelectric properties were also discovered in tilted smectic phases composed of chiral molecules.[3] In subsequent years, antiferroelectric chiral SmC phases[4] and polar properties of achiral bent-core mesogens[5] were also discovered. However, all these liquid crystals are examples of improper ferroelectrics, in which the polar order is a secondary effect. The ordering of dipole moments is induced by steric interactions between the molecules and therefore is usually weak.

For many years, it was believed that dipole-dipole interactions themselves were too weak to produce long-range polar ordering in the liquid state, and the polar order would be disrupted by thermal fluctuations. This common belief was overturned recently by the discovery of the ferroelectric nematic ($N_F$) phase,[6–8] in which the spontaneous electric polarization vector is along the director. In the $N_F$ phase the polar ordering is exceptionally strong, while the viscosity is not much different from that of regular liquids, making these materials interesting for future applications. At first glance, it seems that longitudinal polar order should be even easier to obtain in the smectic phase than in the nematic phase, as thermal fluctuations in lamellar systems are strongly suppressed. The ferroelectric orthogonal smectic ($SmA_F$) phase was first claimed to have been discovered in 1991,[9] but this finding turned out to be premature.[10] Since the discovery of the $N_F$ phase, attention has returned to the search for different ferroelectric smectic phases in combination with the polar $N_F$ phase. This has proved challenging because the requirements for polar order and smectic order are contradictory. Smectic order generally requires molecules to have long tails to enhance the self-segregation of mesogenic cores and alkyl chains that provides the main mechanism of layer formation. However, this inevitably increases the distance between interacting dipoles, which weakens their tendency for order. Therefore, only a limited number of mesogens that show a sequence of polar nematic and smectic phases are known so far.[11–18] In this work, we report the phase behavior and ferroelectric properties of a new liquid crystalline material, which shows the sequence of three polar mesophases: $N_F$, $SmA_F$, and $SmC_F$ below the paraelectric N phase (**Figure 1a**). This allows us, for the first time, to follow the development of polar order and its coupling to positional order and tilt.

The molecular structure of the material studied here is given in **Figure 1** together with its calculated dipole moment, phase sequence, and phase transition temperatures determined from calorimetric studies (**Figure S7**). It shares common fragments with archetypical ferronematogens, RM734,[6] DIO,[7] and UUQU-4N.[19]

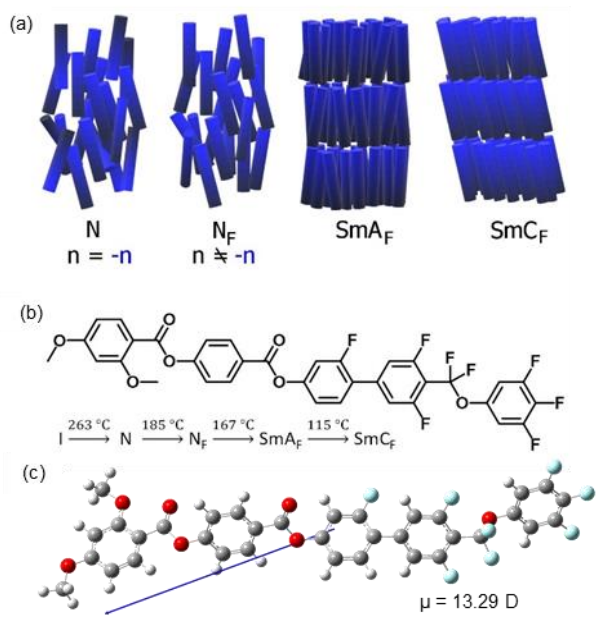

**Figure 1**. (a) Schematic drawing showing the arrangement of polar molecules in N, $N_F$, $SmA_F$ and $SmC_F$ phases. (b) The molecular structure of the mesogen reported here, with the phase transition temperatures. (c) The minimum energy conformation calculated at the B3LYP-GD3BJ/cc-pVTZ level of DFT with arrow showing the direction of the molecular dipole moment.

## 2. Results

### 2.1. Identification of Liquid Crystalline Phases

The material shows a sequence of four liquid crystalline phases. The two higher temperature phases were confirmed to be nematic as only short-range positional order of molecules is observed by X-ray diffraction (XRD). In the two lower temperature phases the low-angle XRD signal narrows to become limited only by instrumental broadening, reflecting the formation of a long-range lamellar structure and indicating that these are both smectic phases (**Figure 2**). In the upper temperature smectic phase, the position of the signal is nearly constant and corresponds to a layer thickness which closely matches the molecular length (30 Å) determined in the crystalline state by single crystal XRD (see SI). This is typical for a smectic A phase. In the lower temperature phase, the layer spacing gradually decreases on lowering the temperature, indicating tilting of molecules within the layers and suggesting that this is a SmC phase. The tilt angle estimated from the change in layer thickness reaches ~15 degrees 30 K below the SmA - SmC phase transition. It should also be noted that in all LC phases the high-angle diffraction signal was diffuse, showing no long-range positional correlations along the short molecular axes, consistent with a sequence of nematic and liquid-like smectic phases.

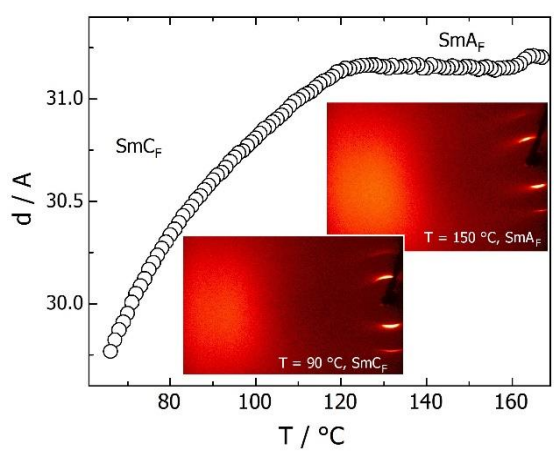

**Figure 2.** Layer spacing vs. temperature; in the inset the 2D XRD patterns registered in $SmA_F$ and $SmC_F$ phases, confirming liquid-like in-plane order of molecules in both phases.

The sequence of LC phases was also followed by observations of characteristic optical textures with polarized-light optical microscopy and measurements of the optical birefringence. In thin (1.5-3 μm) cells with planar anchoring and parallel rubbing on both surfaces both nematic phases and the smectic A phase gave a uniform texture, with the optical axis along the rubbing direction (**Figure 3**). The textures seen for the higher temperature nematic phase are consistent with its identification as a conventional nematic, N, phase. On cooling into the lower temperature nematic phase several conical defects were formed, anchored at the glass pillars that are cell spacers, and such defects are characteristic of the ferroelectric nematic, $N_F$, phase. At the transition from the orthogonal to the tilted smectic phase the uniform texture breaks into small domains, in which the optical axis departs from the rubbing direction. These domains have a characteristic blocky shape with longer sides along the rubbing direction, and in some areas, they also show weak optical activity **(Figure 4).** In optically active domains the molecular orientation at the lower and upper surfaces of the cell differs and inside the cell molecules twist to connect the surface layers. Interestingly, all the liquid crystalline phases showed birefringent textures in cells with homeotropic anchoring (**Figure 5**). The schlieren texture observed in the nematic (N) phase transformed into mosaic-like textures in the lower temperature phases, composed of clearly separated domains. The domains were smooth in the $N_F$ and $SmA_F$ phases and broke into numerous stripes in the $SmC_F$ phase. It should be noted that neither typical focal conic nor fan-shaped textures were observed in the smectic phases, suggesting that polar order is preserved on cooling from the $N_F$ phase.[14]

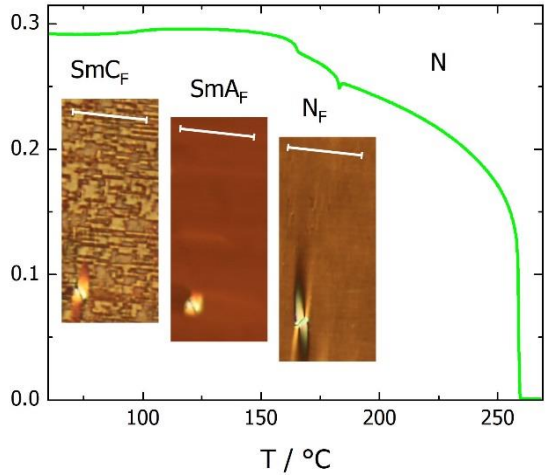

**Figure 3**. Optical birefringence, Δn, vs. temperature. The dip at the N-$N_F$ transition reflects the apparent decrease of the orientational order parameter, S, by ~0.01. The step-like increase of Δn at the $N_F$-$SmA_F$ phase transition corresponds to S changing from 0.78 to 0.81. Inset: the optical textures of $N_F$, $SmA_F$ and $SmC_F$ phases taken in a 1.8-μm-thick cell with planar anchoring and parallel rubbing direction on both surfaces. Scale bars correspond to 50 μm and show the rubbing direction, which is slightly rotated from the polarizer direction.

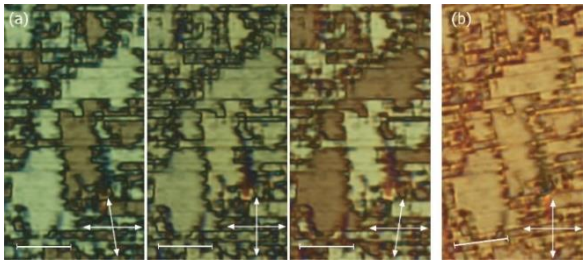

**Figure 4.** Optical texture of the $SmC_F$ phase observed with polarized light microscope in a 1.6-μm-thick cell with planar anchoring. Scale bars correspond to 20 μm and are placed along the rubbing direction. (a) Slight de-crossing of polarizers (arrows) distinguishes domains with opposite optical activity. (b) Upon rotating the sample with respect to crossed polarizers all domains change brightness in the same manner. This indicates that direction of the optical axis is the same in the different domains and thus the domains are not defined by different tilt direction.

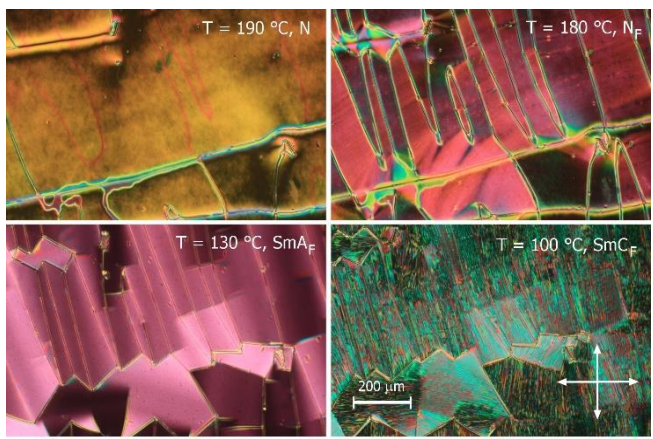

**Figure 5.** Optical textures observed with polarized light microscopy in LC phases of the reported compound in a 5-μm-thick cell with homeotropic anchoring. Arrows indicate the polarizers direction.

The optical birefringence, $\Delta n$, increases continuously in the nematic phase, following a critical, power-law dependence of the orientational order of the molecules (**Figure 3**). The trend continues also in the $N_F$ phase, however close to the N-$N_F$ transition a pronounced dip in $\Delta n$ is observed. It appears that the development of polar order is accompanied by strong orientational order fluctuations (splay deformations), lowering the effective optical anisotropy.[20] The $N_F$-$SmA_F$ phase transition is marked by a step-like increase of birefringence, indicating that the formation of long-range positional order is accompanied by a small increase in the orientational order of molecules; the order parameter S jumps from 0.78 to 0.81, and such a discontinuity is often observed at weakly first-order transitions. At the $SmA_F$-$SmC_F$ phase transition, the measured $\Delta n$ values decrease, however this apparent behavior may be ascribed to distortion of the uniform texture as described above.

Calorimetric studies revealed that three of the observed LC phases are enantiotropic, the pristine crystal melts at 145 °C into the $SmA_F$ phase, and the monotropic $SmC_F$ phase can be observed due to the supercooling effect (**Figure S7**). The $N_F$-N phase transition was accompanied by a jump in the heat capacity, characteristic of a second-order phase transition, while at the $SmA_F$ - $N_F$ phase transition a small enthalpy peak was registered, which is characteristic for a weakly first-order transition. These findings are consistent with the optical studies. Despite the clearly visible optical texture changes at the $SmA_F$-$SmC_F$ phase transition, no change in the heat capacity was detected in calorimetric measurements, suggesting that the phase transition is second order with a change in heat capacity below the detection limit of the DSC.

From these observations, we can see that the studied material shows a sequence of liquid crystal phases with gradually increasing molecular order: nematic → orthogonal smectic → tilted smectic. Taking into account the strong molecular dipole moment, the polar properties of the LC phases should also be considered. While the ferroelectric nature of the lower temperature nematic phase was indicated by the observed optical textures characteristic of the $N_F$ phase, the properties of the smectic phases require further investigation, and this will be discussed in the following section.

## 2.2. Polar Nature of the Liquid Crystal Phases

The studied molecule is highly polar, and its dipole moment was calculated to be 13.29 D (using DFT at the B3LYP-GD3BJ/cc-pVTZ level). While it must be remembered that this calculation considers only a single conformation of an isolated molecule in the gas phase, and hence may be a slight overestimation of the true value of the dipole moment for an average conformation in the LC phases, the calculated value is consistent with those reported for similar ferroelectric LCs, and much greater than those of conventional mesogenic materials, e.g. ca. 4 D for 5CB.[21] The dipole moment in the studied molecule is offset only slightly from the long molecular axis (**Figure 1**), as reported for similar molecules.[11,16] Therefore, in polar phases, in which molecules rotate freely around long axis the spontaneous electric polarization vector is expected to lie along the director.

It should be noted that all the LC phases below the N phase are SHG active at zero electric field (**Figure 6(a)** and **(b)**), which confirms the ferroelectric character of their ground state (SmA$_F$ and SmC$_F$). Their response to an applied electric field has been studied by observation of optical texture changes in cells with transparent electrodes. The application of an electric field across the cell thickness in the N$_F$ and SmA$_F$ phases induces a homeotropic texture as the polarization, and thus director, aligns along the electric field (**Figure S8**). In the N$_F$ phase, when the field is reduced to zero the sample relaxes immediately to a birefringent state with polarization inclined to the cell surface to reduce surface bound charges.[18] In the SmA$_F$ phase the relaxation to a birefringent texture takes place but the process is slow (taking several seconds) as the reorientation of polarization requires layer rotation. In the SmC$_F$ phase, applying an electric field also leads to a non-birefringent state, however, this occurs through an intermediate state with reduced birefringence (**Figure S8**). This suggests a two-step switching mechanism, involving reversal of polarization through reorientation of the layers, and in addition, a second process related to the removal or restoration of the director tilt.

The ferroelectric nature of the smectic phases has been further confirmed by observation of a clear switching current peak when ac voltage is applied across the sample (**Figure S9**). The spontaneous electric polarization, calculated from the peak area, increases in the N$_F$ phase and reaches ~5 µC cm$^{-2}$ in the smectic phases. Moreover, the ferroelectric ground state of the smectic A$_F$ phase was confirmed by studying the switching behavior using a modified triangular wave voltage (**Figure 6(a)**). In this experiment two successive positive voltage pulses (separated by period with zero voltage and followed by two negative ones) were applied and the switching current peak was observed only for the first pulse of each sequence. This indicates that reducing the voltage from the maximum value to zero preserves the ferroelectric arrangement of dipoles.

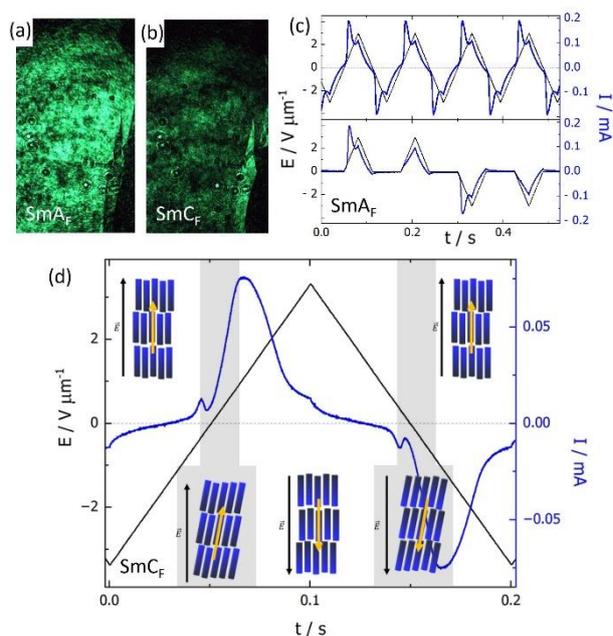

**Figure 6.** Images taken with SHG microscope in ground state (no applied voltage) of (a) SmA$_F$ and (b) SmC$_F$ phases for a sample prepared in a cell with planar anchoring. Incident IR ($\lambda$=1064 nm) radiation resulted in emission of green (SH) light in both phases, proving their ferroelectric properties. (c) Switching current in the SmA$_F$ phase (blue lines) under application of triangular and modified triangular wave voltage (black lines). Application of two successive pulses of the same sign results in the current peak appearing only in the first pulse, confirming the stable ferroelectric ground state of the phase. (d) The switching current in the SmC$_F$ phase (blue lines) under application of triangular wave voltage (black lines), showing the main switching peak and the additional smaller peak at low voltages. A schematic representation of the proposed two-step switching process in the SmC$_F$ phase is given, with the black arrow representing the direction of the applied electric field and the yellow arrow corresponding to the direction of the spontaneous polarization.

In the SmC$_F$ phase, under application of triangular-wave voltage, apart from the main switching current peak due to reversal of polarization, an additional small peak is detected, positioned close to 0V (**Figure 6(d)**). Such a peak has been observed in other recently reported polar SmC phases.[12,22] This small peak decreases on heating and finally disappears on entering the SmA$_F$ phase (**Figure S9**). The underlying mechanism producing this signal must be a very low energy process. One possibility may be that this corresponds to the reappearance of the molecular tilt within the SmC$_F$ phase, and a sketch of such a possibility is given in **Figure 6(d)**. The whole switching sequence in the SmC$_F$ phase would involve three steps: first, the polarization within the layers aligns with the electric field when a high voltage is applied – in this state the tilt angle is removed and the phase becomes orthogonal; secondly, when the electric field is reduced, molecular tilt is regained, which changes the electric polarization along the layer normal and causes the small current peak; and finally, after reversal and increase of the applied voltage the polarization switches and realigns with the applied electric field, giving rise to the main recorded current peak.

Finally, dielectric spectroscopy studies were conducted to follow the characteristic fluctuations of polar order in the LC phases. Although the interpretation of dielectric measurements for strongly polar phases is not straightforward,[23,24] and measured values of both the position and dielectric strength of relaxation modes are influenced by the thickness and type of the cells used, the relative changes in dielectric constant largely reflect the material properties. The studied compound was examined in a 10-μm-thick cell with gold electrodes and no polymer aligning layers, to avoid contribution from the polymer layer capacitance. In such cells the orientation of the director with respect to the measuring electric field is random. In the $N_F$ phase, a strong, nearly temperature-independent dielectric response was detected, with a relaxation frequency of ~$10^4$ Hz (**Figure 7**). In the $SmA_F$ phase, a much lower permittivity is measured in the whole tested frequency range, 10-$10^7$ Hz. On lowering the temperature and upon approaching the tilted smectic phase, a weak, high frequency mode starts to build in the $SmA_F$ phase, with a relaxation frequency that critically decreases, and a mode strength that critically increases; such behavior can be ascribed to the softening of the tilt angle fluctuations. In the $SmC_F$ phase the soft mode condenses and a strong dielectric response is restored, with relaxation at ~$10^4$ Hz. The low permittivity in the ferroelectric $SmA_F$ phase in comparison to the strong response in the $SmC_F$ phase might be explained by the different types of polar fluctuations possible in these phases. In the non-tilted smectic $A_F$ phase, two basic fluctuation modes can be considered: changes in the magnitude of the polarization vector and undulation of the polar layers. The first mechanism is active only when the system is in the close vicinity of a transition to a paraelectric phase, which is not the case in the material studied here. The second mechanism involves layer undulations, which should be strongly suppressed in highly polar systems as they produce bound charges related to local splay of the polarization vector. In the $SmC_F$ phase, an additional fluctuation mechanism is activated due to molecular tilt – the collective rotation of molecules on the tilt cone. Due to the lower energy required for such fluctuations, the strong dielectric response is restored in this phase.

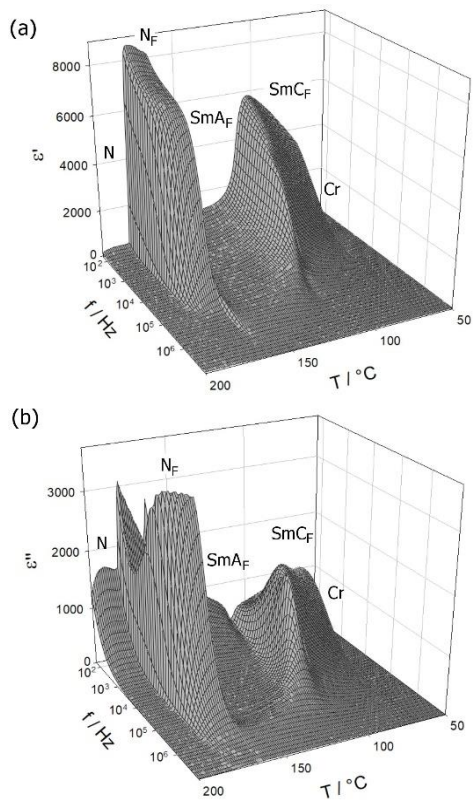

**Figure 7.** Real (a) and imaginary (b) parts of apparent dielectric permittivity measured in a 10-μm-thick cell with gold electrodes.

Combining the insights from the measurements described above and the properties of the three polar phases studied, we can consider the relation between polar and orientational order in these phases. In all three phases, the dipole moment is expected to follow, approximately, the director of the LC phase. As such, in the orthogonal smectic $A_F$ phase the polarization is along the layer normal, while it is inclined by a tilt angle in the $SmC_F$ phase. Such an arrangement allows the development of lamellar order without the need for any dramatic reorganization of the polar structure between the different LC phases, allowing for the observed sequence of ferroelectric phases without intermediate antiferroelectric or paraelectric phases.

## 3. Conclusion

In summary, the material studied here shows a sequence of three ferroelectric phases: $N_F$-$SmA_F$-$SmC_F$, and our results show that the development of lamellar structure has only a minor effect on the polar order. The polar, ferroelectric character is uninterrupted across the $SmC_F$, $SmA_F$ and $N_F$ phases. While the $N_F$-$SmA_F$ phase transition is weakly first order, the $SmA_F$-$SmC_F$ transition is second order. Although the polarization value determined from the switching current is similar in both ferroelectric smectic phases, their dielectric response is very different. In the orthogonal ferroelectric smectic phase polar fluctuations are strongly suppressed, while in the tilted $SmC_F$ phase collective rotation of molecules on a tilt cone give rise to a strong dielectric response. Both smectic phases easily respond to an electric field; in the $SmA_F$ phase under a reversing electric field switching takes place between the two optically homeotropic states, while in the $SmC_F$ phase an intermediate state with schlieren texture is formed with gradually reduced birefringence, showing that under an electric field the cone angle is reduced to zero.

## 4. Methods

The full synthetic and chemical characterization details are described in the SI, as well as molecular structure and crystal structure parameters obtained from single crystal X-ray diffraction (XRD) experiment (**Figure S5-S7**).

[CCDC **2375621** contains the supplementary crystallographic data for this paper. These data can be obtained free of charge from The Cambridge Crystallographic Data Centre via www.ccdc.cam.ac.uk/data_request/cif.]

**Supporting Information**

Supporting Information is placed after thin manuscript.

**Acknowledgements**

This research was supported by the National Science Centre (Poland) under the grant no. 2021/43/B/ST5/00240. We gratefully acknowledge Polish high-performance computing infrastructure PLGrid (HPC Center: ACK Cyfronet AGH) for providing computer facilities and support within computational grant no. PLG/2024/017463.

Supporting Information for

**Nematic and Smectic Phases with Proper Ferroelectric Order**

Grant J. Strachan, Ewa Górecka, Jadwiga Szydłowska, Anna Makal, Damian Pociecha

**Experimental Methods**

Transition temperatures and the associated enthalpy changes were measured by differential scanning calorimetry using a TA DSC Q200 instrument. Measurements were performed under a nitrogen atmosphere with a heating/cooling rate of 10 K min$^{-1}$, unless otherwise specified.
Observations of optical textures of liquid crystalline phases was carried out by polarised-light optical microscopy using a Zeiss AxioImager.A2m microscope equipped with a Linkam heating stage.
Optical birefringence was measured with a setup based on a photoelastic modulator (PEM-90, Hinds) working at a modulation frequency $f = 50$ kHz; as a light source a halogen lamp (Hamamatsu LC8) equipped with narrow bandpass filters was used. The transmitted light intensity was monitored with a photodiode (FLC Electronics PIN-20) and the signal was deconvoluted with a lock-in amplifier (EG&G 7265) into 1$f$ and 2$f$ components to yield a retardation induced by the sample. Knowing the sample thickness, the retardation was recalculated into optical birefringence. Samples were prepared in 1.6-μm-thick cells with planar anchoring. The alignment quality was checked prior to measurement by inspection under the polarised-light optical microscope.
X-ray diffraction measurements of samples in liquid crystalline phases were carried out using a Bruker D8 GADDS system, equipped with micro-focus-type X-ray source with Cu anode and dedicated optics and VANTEC2000 area detector.  Small angle diffraction experiments were performed on a Bruker Nanostar system (IμS microfocus source with copper target, MRI heating stage, Vantec 2000 area detector).

Single-crystal X-ray diffraction data were collected on a SuperNova diffractometer with micro-focus sealed source of MoKα X-ray radiation (λ = 0.71073 Å) and CCD Eos detector. Single crystals of studied compound were obtained from a chloroform solution using hexane as an antisolvent. A suitable crystal – a colorless prism - was mounted on a nylon loop with a trace of ParatoneN oil. The crystal was kept at 120.00(10) K during data collection in cold nitrogen stream using Oxford Cryosystems device. Data reduction was performed with CrysAlisPro.[1] Gaussian absorption correction was applied using spherical harmonics with SCALE3 ABSPACK algorithm.  Using Olex2,[2] the structure was solved with the olex2.solve[3] program using Charge Flipping and refined with the olex2.refine[3] package using Gauss-Newton minimization. H-atom positions were identifiable from a difference Fourier map but were refined with distances restrained to standardized values and the atomic displacement parameters (ADP-s) of H atoms were restrained as 'riding' on the displacement parameters of the covalently bound non-H atoms. A static disorder concerning the position of the fluorine F8 was refined, yielding 93% of the major component (F8 bound to C19) and 7% of the minor component with F8a bound to C15. Similarity restraints were applied for C – F distances and F displacement parameters of both disorder components.

Spontaneous electric polarisation was determined by integration of the current peaks recorded during polarization switching upon applying a triangular-wave voltage. 3- to 10-μm-thick cells with ITO or gold electrodes and no polymer aligning layers were used, and the switching current was determined by recording the voltage drop on a resistor connected in series with the sample.

The SHG response was investigated using a microscopic setup based on a solid-state laser EKSPLA NL202. Laser pulses (9 ns) at a 10 Hz repetition rate and max. 2 mJ pulse energy at λ=1064 nm were applied. The pulse energy was adjusted for each sample to avoid its decomposition. The infra-red beam was incident onto a LC homogenous cell of thickness 5 μm. An IR pass filter was placed at the entrance to the sample and a green pass filter at the exit of the sample.

The complex dielectric permittivity, ε*, was measured using a Solartron 1260 impedance analyser, in the 1 Hz −10 MHz frequency range, and a probe voltage of 50 mV. The material was placed in a 5- or 10-μm-thick glass cell with gold electrodes. Cells without polymer aligning layers were used, as the presence of the thin (~10 nm) polyimide layers at the cell surfaces acts as an additional high capacitance capacitor in a series circuit with the capacitor filled with the LC sample, which for materials with very high values of permittivity, may strongly affect the measured permittivity of the LC phases. Lack of a surfactant layer resulted in a random configuration of the director in the LC phases.

DFT geometry optimization was carried out at the B3LYP- GD3BJ/cc-pVTZ level of theory using Gaussian 16 (Revision C.01)[4] on the Ares cluster of the Polish high-performance computing infrastructure PLGrid (HPC Center: ACK Cyfronet AGH. Following geometry optimization, a frequency calculation was used to confirm that the obtained structure was at an energy minimum.

**Synthetic Procedures and Structural Characterisation**

Unless otherwise stated, all materials were obtained from commercial sources and used without further purification.
Reactions were monitored using thin layer chromatography (TLC) using aluminium-backed plates with a coating of Merck Kieselgel 60 F254 silica and an appropriate solvent system. Spots were visualised using UV light (254 nm). Flash column chromatography was carried out using silica grade 60 Å 40-63 micron.
FT-IR spectra were obtained using a Nicolet iS50 FT-IR spectrometer. $^1$H, $^{19}$F, and $^{13}$C NMR spectra were recorded on a 400 MHz Agilent NMR spectrometer using either CDCl$_3$ or DMSO-$d_6$ as solvent and using residual non-deuterated trace solvents as reference. Chemical shifts (δ) are given in ppm relative to TMS (δ = 0.00 ppm). Mass spectroscopy was conducted on a Micromass LCT instrument.

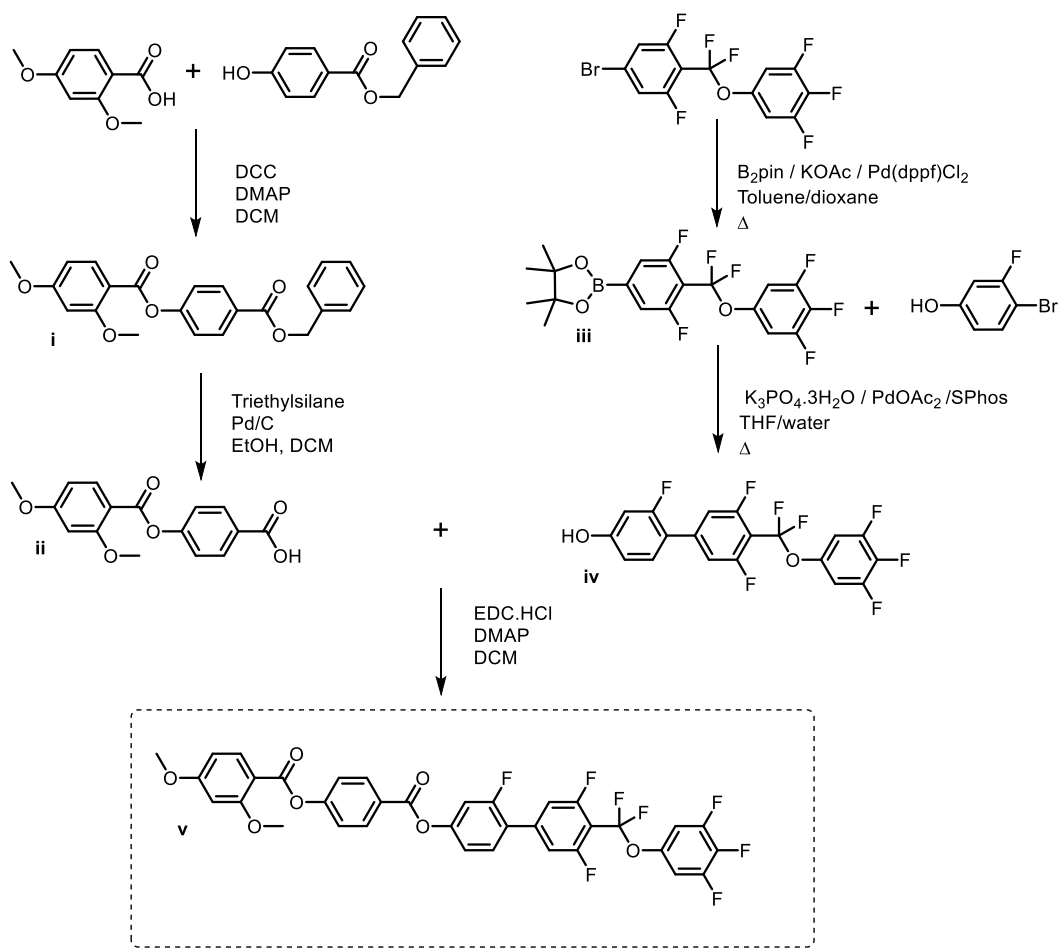

*Scheme 1: Synthetic route to the new material reported here.*

Compounds **i**,[5] **ii**,[5] **iii**,[6] and **iv**[7] have been previously reported.

**Benzyl ester i**
2,4-dimethoxy benzoic acid (1.190 g, 6.6 mmol, 1.1 eq) and *N,N'*-dicyclohexylcarbodiimide (DCC) (1.730 g, 8.4 mmol, 1.4 eq) were dissolved in DCM (50 ml) and stirred for 10 minutes. Benzyl 4-hydroxybenzoate (1.371 g, 6 mmol, 1 eq) and 4-dimethylaminopyridine (DMAP) (74 mg, 0.6 mmol, 0.1 eq) were added and the reaction was left stirring at room temperature overnight. The reaction was filtered to remove the dicyclohexylurea and the solvent removed *in vacuo*. The crude product was recrystallised from ethanol to yield the product as a white solid. (1.27 g, 54 %).
$^1$H NMR (400 MHz, CDCl$_3$) δ 8.13 (d, $J$ = 8.6 Hz, 2H), 8.07 (d, $J$ = 8.7 Hz, 1H), 7.39 (overlapping multiplets, 5H), 7.28 (d, $J$ = 8.6 Hz, 2H), 6.56 (dd, $J$ = 8.7, 2.2 Hz, 1H), 6.53 (d, $J$ = 2.2 Hz, 1H), 5.37 (s, 2H), 3.91 (s, 3H), 3.89 (s, 3H). $^{13}$C NMR (101 MHz, CDCl$_3$) δ 165.81, 165.20, 162.95, 162.39, 154.97, 135.99, 134.58, 131.17 (2C), 128.60 (2C), 128.25, 128.17 (2C), 127.25, 122.02 (2C), 110.61, 104.86, 98.98, 66.74, 56.02, 55.60.

**Acid ii**
Under an argon atmosphere triethylsilane (1.6 ml, 10 mmol, 10 eq.) was added dropwise to a stirred solution of **i** (393 mg, 1 mmol) and 5 % Pd/C (80 mg) in ethanol (3 ml) and DCM (3 ml). The reaction was stirred for 5 minutes after addition was complete, then filtered through celite and the solvent removed *in vacuo*. The crude product was washed with hexane to yield the product as a white powder. (300 mg, quant.) R$_f$ 0.22 (DCM)

¹H NMR (400 MHz, CDCl₃) δ 8.15 (d, *J* = 8.5 Hz, 2H), 8.08 (d, *J* = 8.8 Hz, 1H), 7.32 (d, *J* = 8.5 Hz, 2H), 6.57 (dd, *J* = 8.8, 2.2 Hz, 1H), 6.54 (d, *J* = 2.2 Hz, 1H), 3.93 (s, 3H), 3.90 (s, 3H). ¹³C NMR (101 MHz, DMSO-$d_6$) δ 167.16, 165.35, 163.00, 162.15, 154.67, 134.44, 131.27 (2C), 128.70, 122.70 (2C), 110.32, 106.14, 99.43, 56.43, 56.16.

**Boronic ester iii**
A solution of 5-bromo-2-(difluoro(3,4,5-trifluorophenoxy)methyl)-1,3-difluorobenzene (4.951 g 12.7 mmol), bis(pinacolato)diboron (3.35 g 12.7 mmol), anhydrous potassium acetate (3.67 g 36.7 mmol) in toluene (100 ml) and 1,4-dioxane (100 ml) was sparged with argon for 1 hr. Then the catalyst Pd(dppf)Cl₂ (280 mg, 3 mol%) was added and the reaction was stirred at 80°C for 5 hrs. The reaction was cooled to room temperature and added to 1M HCl solution. The product was extracted with toluene and the organic layer was washed three times with water and dried over MgSO₄. The solvent was removed under vacuum and the crude solid thus obtained was recrystallized from ethanol. Yield 2.5 g, 45 %).
¹H NMR (400 MHz, CDCl₃) δ 7.38 (d, *J* = 10.5 Hz, 2H), 6.96 (m, 2H), 1.35 (s, 12H). ¹³C NMR (101 MHz, CDCl₃) δ 160.81, 160.78, 160.78, 158.23, 158.20, 152.26, 152.26, 152.20, 152.15, 152.15, 152.10, 149.76, 149.76, 149.71, 149.65, 149.65, 149.60, 144.72, 144.61, 144.58, 144.52, 144.47, 139.79, 139.63, 139.63, 139.48, 137.29, 137.14, 136.99, 122.83, 120.18, 120.18, 118.17, 118.14, 118.12, 118.12, 117.96, 117.95, 117.92, 117.92, 111.65, 111.42, 111.19, 107.52, 107.45, 107.35, 107.28, 84.86, 24.80.

**Phenol iv**
Intermediate **iii**, (1.108 g, 2.5 mmol) 4-bromo-3-fluorophenol, (0.587 g, 2.7 mmol) and potassium phosphate trihydrate (2.32 g, 8.7 mmol) were dissolved in THF (20 ml) and distilled water (1 ml), sparged with argon, and refluxed for 1 hour. Palladium acetate (33 mg, 0.15 mmol) and S-Phos (101 mg, 0.25 mmol) were added, and the reaction was refluxed for 4 hours. The mixture was cooled to RT, acidified with 1 M HCl, extracted with three portions of DCM and the organic layers were combined, dried over magnesium sulfate and the solvent removed *in vacuo*. The crude product was recrystallised from hexane to yield the desired product. (404 mg, 38 %).
¹H NMR (400 MHz, CDCl₃) δ 7.31 (t, *J* = 8.6 Hz, 1H), 7.16 (d, *J* = 11.0 Hz, 2H), 6.99 (m, 2H), 6.72 (m, 2H), 5.39 (s, 1H). ¹³C NMR (101 MHz, CDCl₃) δ 161.55, 161.16, 161.10, 159.05, 158.62, 158.60, 158.54, 157.86, 157.74, 152.28, 152.23, 152.17, 152.12, 149.79, 149.74, 149.68, 149.63, 141.51, 141.40, 141.30, 139.80, 139.65, 139.50, 137.31, 137.16, 137.00, 130.84, 130.79, 120.22, 118.06, 117.94, 112.76, 112.73, 112.70, 112.52, 112.49, 112.45, 112.16, 112.13, 107.55, 107.48, 107.37, 107.31, 104.21, 103.96. ¹⁹F NMR (376 MHz, CDCl₃) δ -61.72 (t, *J* = 26.3 Hz, 2F), -111.03 (td, *J* = 26.3, 10.3 Hz, 2F), -114.51 (t, *J* = 10.3 Hz, 1F), -132.53 (dd, *J* = 20.9, 8.0 Hz, 2F), -163.22 (tt, *J* = 20.9, 5.8 Hz. 1F).

**Target ester v**
Intermediate acid **iii** (71 mg, 0.24 mmol) and EDC.HCl (62 mg, 0.32 mmol) were dissolved in 10 ml DCM and stirred for 5 minutes. Phenol **iv** (90 mg, 0.21 mmol) and DMAP (3 mg, 0.02 mmol) were added, and the mixture stirred overnight and monitored by TLC ($R_f$ 0.58 DCM). The reaction mixture was washed 3 times with water and the organic layer dried over magnesium sulfate and removed *in vacuo*. The crude solid was purified by column chromatography (gradient elution, 50/50 DCM: hexane → DCM) and triturated with hexane. Yield 63 mg (42 %).
¹H NMR (400 MHz, CDCl₃) δ 8.26 (d, *J* = 8.6 Hz, 2H), 8.10 (d, *J* = 8.7 Hz, 1H), 7.50 (t, *J* = 8.5 Hz, 1H), 7.39 (d, *J* = 8.5 Hz, 2H), 7.27 – 7.14 (overlapping multiplets and chloroform, 5H), 7.04 – 6.97 (m, 2H), 6.58 (d, *J* = 9.1 Hz, 1H), 6.55 (s, 1H), 3.94 (s, 3H), 3.91 (s, 3H). ¹³C NMR (101 MHz, CDCl₃) δ 165.35, 163.97, 162.80, 162.50, 161.22, 161.16, 160.78, 158.65, 158.59,

158.27, 155.83, 152.39, 152.30, 152.28, 152.25, 152.20, 152.15, 149.81, 149.76, 149.70, 149.65, 140.90, 140.87, 140.77, 140.77, 140.67, 140.64, 139.84, 139.71, 139.69, 139.54, 137.35, 137.20, 137.05, 134.63, 131.82, 130.56, 130.52, 125.77, 123.28, 123.25, 123.13, 122.44, 120.11, 120.08, 118.50, 118.47, 113.24, 113.21, 113.18, 113.00, 112.97, 112.93, 111.07, 110.81, 110.35, 107.58, 107.52, 107.41, 107.34, 104.94, 98.98, 56.04, 55.62. $^{19}$F NMR (376 MHz, CDCl$_3$) δ -61.82 (t, $J$ = 26.4 Hz, 2F), -110.45 (td, $J$ = 26.4, 10.4 Hz, 2F), -113.83 – -113.93 (m, 1F), -132.37 – -132.55 (m, 2F), -163.12 (tt, $J$ = 20.9, 5.9 Hz, 1F).

HRMS (ESI) m/z Calculated for C$_{35}$H$_{20}$O$_7$F$_8$:
[M+H]$^+$ theoretical mass: 705.11540, found 705.11657, difference 1.65 ppm.
[M+Na]$^+$ theoretical mass: 727.09735, found 727.09838, difference 1.42 ppm.
IR (ν$_{max}$/cm$^{-1}$) 3101, 3074, 3005, 2926, 2847, 1741, 1710, 1611, 1519, 1233, 1135, 1037.

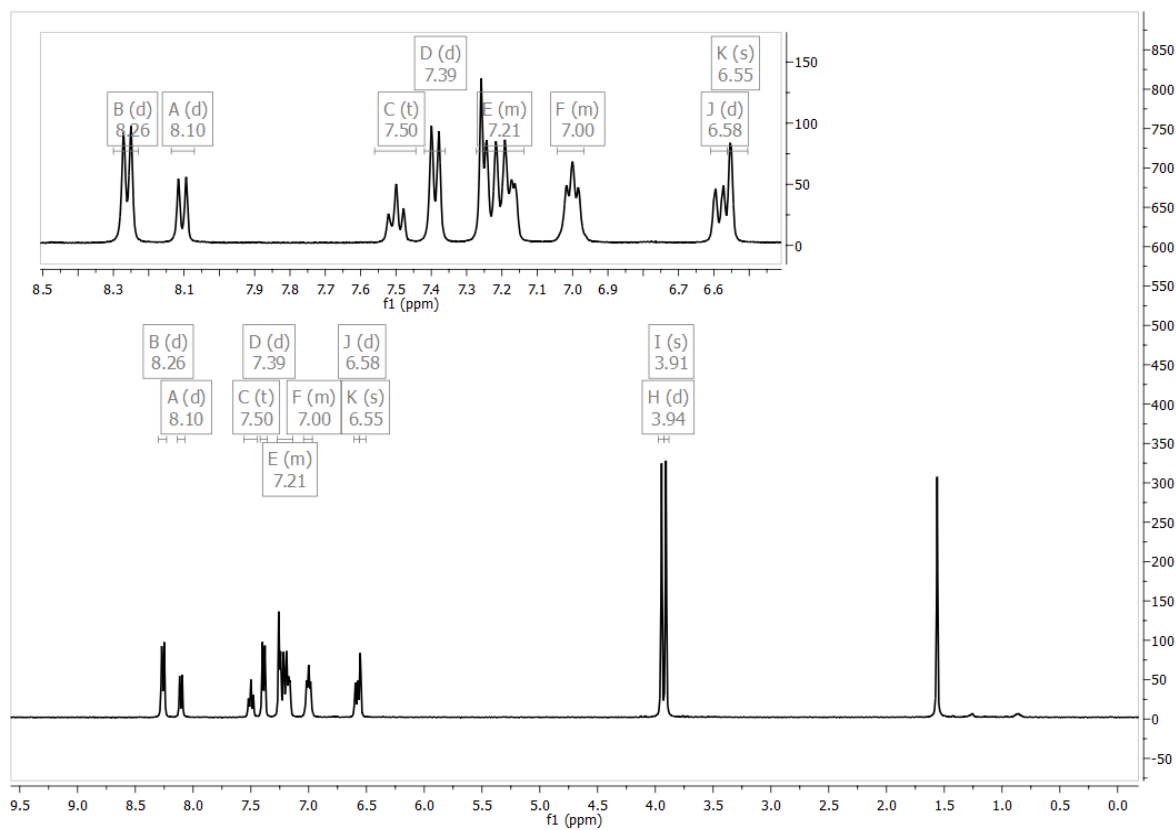

Figure S1: $^1$H NMR spectrum of compound **v** in CDCl$_3$.

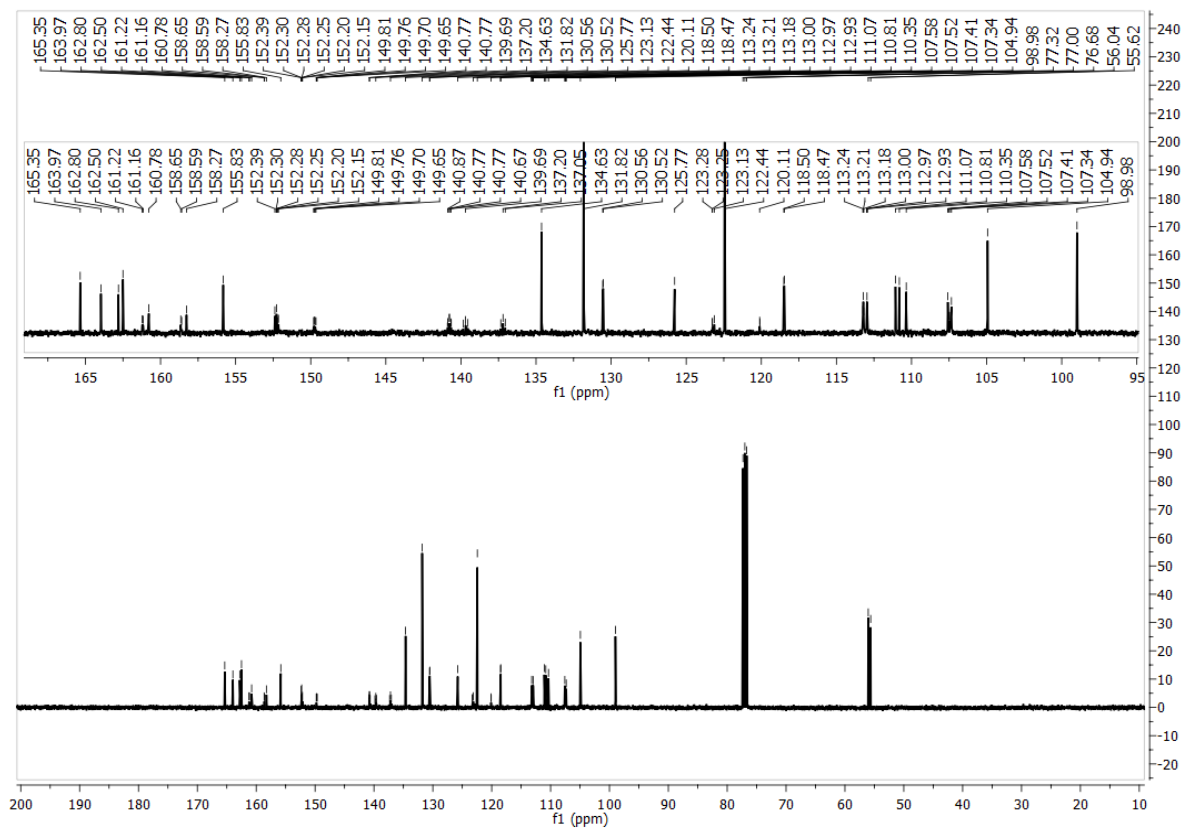

*Figure S2 $^{13}$C NMR spectrum of compound **v** in CDCl$_3$.*

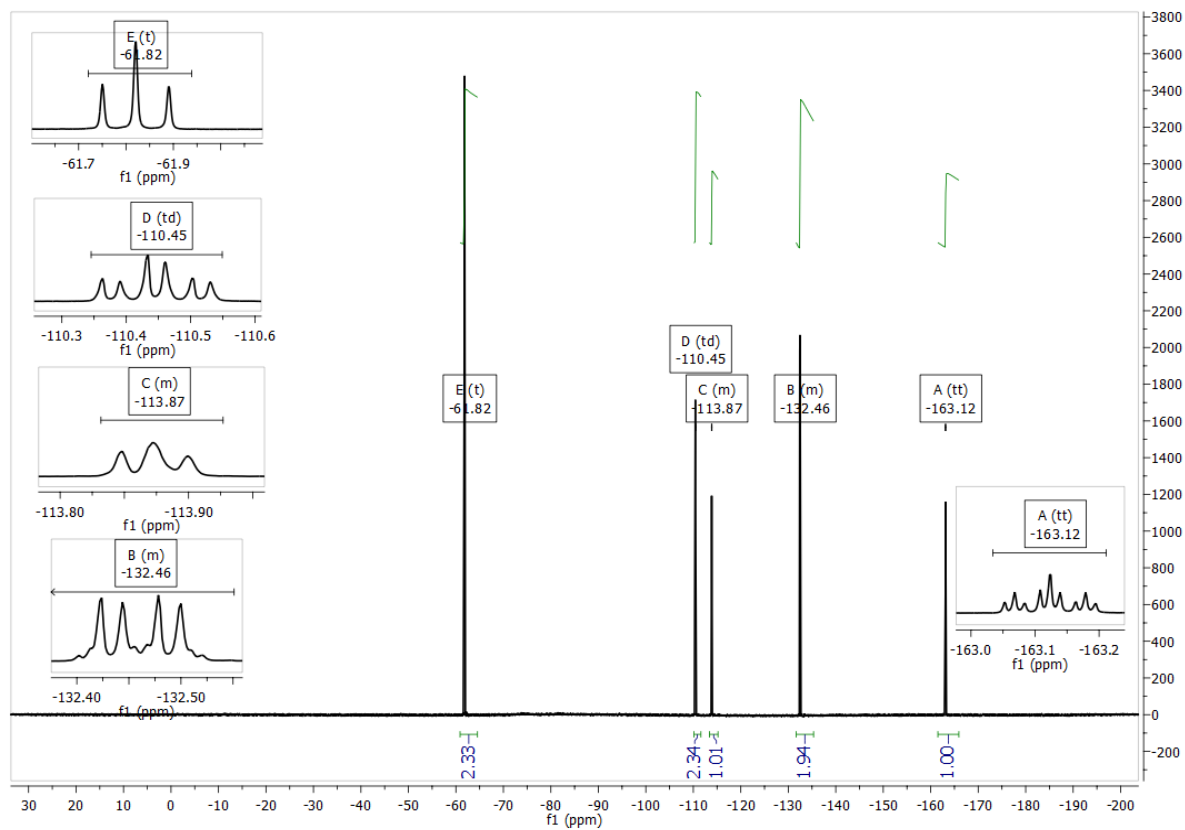

*Figure S3: $^{19}$F NMR spectrum of compound **v** in CDCl$_3$.*

## Single crystal X-ray diffraction

The material crystallized in centrosymmetric P-1 space group in triclinic system, with one independent molecule in the asymmetric unit. There were no traces of the solvent molecules in voids in this crystal structure. There is a static disorder concerning the position of the fluorine F8 with 93% of the major component with F8 bound to C19 and 7% of the minor component (F8a bound to C15, Figure S4).

The molecule in the crystal structure is fully stretched with no co-planar phenyl rings. The terminal rings #1 and #5 are nearly perpendicular to the closest-connected rings #2 and #4 accordingly, while the three middle rings are rotated at about 40° with respect to each other (Fig. S4 and Table S1), in order to avoid short F … F and F … H intramolecular contacts. The ester groups are also not co-planar with the attached phenyl rings: O2 and O4 atoms are out of plane of the ring #4 and #5 accordingly by 24.83(17)° and 23.45(17).

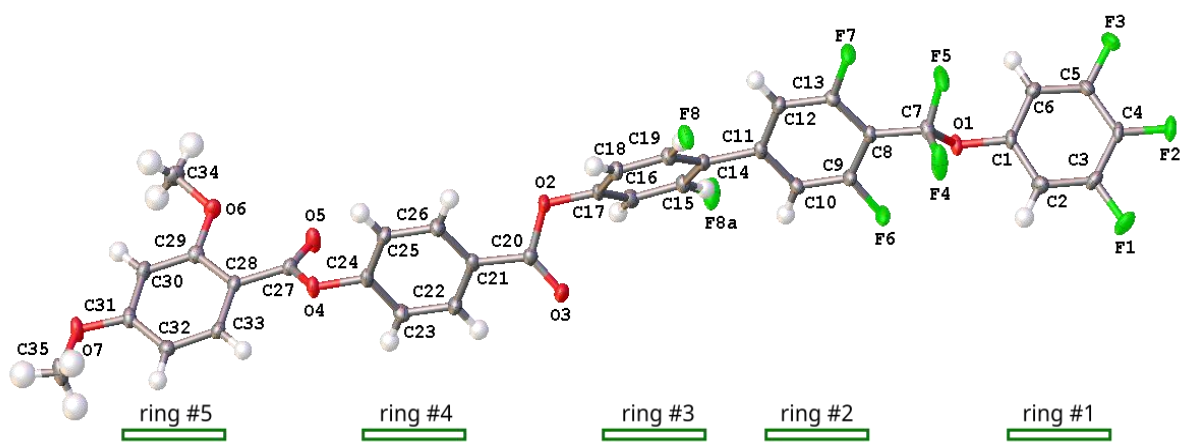

*Figure S4. Determined structure of molecule with the applied numbering scheme. H atom numbers are the same as those of the closest covalently bound C atom. Atomic displacement parameters represented at 50% probability level. The minor variant of disorder labelled as F8a. Aromatic rings in the structure have been assigned numbers based on the order of C atoms.*

*Table S1. Angles between the planes of the phenyl rings.*

| Phenyl rings | Angle [°] |
|---|---|
| #1 - #2 | 126.81(6) |
| #2 - #3 | 39.91(6) |
| #3 - #4 | 36.53(6) |
| #4 - #5 | 107.36(6) |

The most important intermolecular interactions have been identified using interaction energy estimation with UNI potential[8,9] within Mercury[10] and illustrated in Figure S5. Notably, the two strongest interactions involve intermolecular π … π stacking of the ring #5 with its symmetry-related equivalent and #2 with #4, while the third occurs between the molecules related by translation and appears to arise from maximizing the number of C – H … F interactions (Tables S2, S3).

As a consequence, the molecules all align approximately with the crystallographic [1-1-1] direction but do not form separate columns or layers in the crystal structure (Figure S6).

*Table S2. Geometry of intermolecular π ... π interactions in the crystal structure.*

| phenyl rings involved | inter-planar angle [°] | inter-centroid distance [Å] | lateral shift [Å] |
|---|---|---|---|
| #1---#1(-1+X,1+Y,+Z) | 3.422(6) | 3.877(6) | 1.773(6) |
| #2---#4(-X,1-Y,1-Z) | 0.000 | 3.745(7) | 1.786(7) |
| #5---#5(4-X,-2-Y,-Z) | 0.000 | 3.514(6) | 1.102(7) |

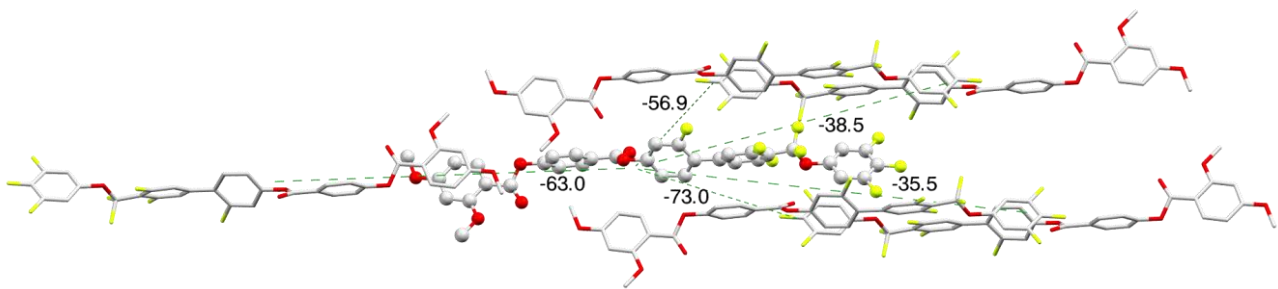

*Figure S5. The most important intermolecular interactions in the crystal structure represented as dashed lines between the centers of interacting molecules. Among the three strongest interactions, two rely on π ... π stacking.*

*Figure S6. Visualization of crystal packing in selected directions: (a) view along [100], (b) view along [010] and (c) view along [1-1-1]. Phenyl rings outlined in brown.*

*Table S3. The strongest intermolecular interactions.*

| | Intermolecular Distance [Å] | Energy [ kJ/mol] | |
|---|---|---|---|
| 1 | 11.16 | -72.9 | π … π interactions (ring #5) |
| 2 | 20.83 | -63.0 | π … π interactions (rings #2 and #4) |
| 3 | 7.82 | -56.8 | C – H … F and C – H … π |
| 4 | 19.78 | -38.5 | C – H … F and C – H … π |
| 5 | 24.16 | -35.5 | π … π interactions (ring #1) |

The final crystal structure was deposited with the CCDC (**CCDC Number 2375621**).

Detailed information on X-ray data and crystal structure model are summarized in Table S4.

*Table S4. Crystal data and structure refinement for the studied compound*

| | |
|---|---|
| Empirical formula | $C_{35}H_{20}F_8O_7$ |
| CCDC code | 2375621 |
| Formula weight | 704.532 |
| Temperature/K | 120.00(10) |
| Crystal system | triclinic |
| Space group | P-1 |
| a/Å | 7.8233(3) |
| b/Å | 9.3769(4) |
| c/Å | 20.2777(8) |
| α/° | 83.455(3) |
| β/° | 89.797(3) |
| γ/° | 80.365(3) |
| Volume/Å$^3$ | 1456.82(10) |
| Z | 2 |
| ρ$_{calc}$g/cm$^3$ | 1.606 |
| μ/mm$^{-1}$ | 0.145 |
| F(000) | 716.7 |
| Crystal size/mm$^3$ | 0.211 × 0.098 × 0.037 |
| Radiation | Mo Kα (λ = 0.71073) |
| 2Θ range for data collection/° | 4.04 to 58.9 |
| Index ranges | -10 ≤ h ≤ 10, -12 ≤ k ≤ 12, -26 ≤ l ≤ 27 |
| Reflections collected | 21214 |
| Independent reflections | 7454 [R$_{int}$ = 0.0426, R$_{sigma}$ = 0.0477] |

| | | | | |
|---|---|---|---|---|
| Data/restraints/parameters | | 7454/1/463 | | |
| Goodness-of-fit on $F^2$ | | 1.056 | | |
| Final R indexes [I>=2σ (I)] | | $R_1$ = 0.0517, $wR_2$ = 0.1078 | | |
| Final R indexes [all data] | | $R_1$ = 0.0700, $wR_2$ = 0.1165 | | |
| Largest diff. peak/hole / e Å$^{-3}$ | | 0.46/-0.34 | | |

Intermolecular energy estimation was based on an empiric potential,[8,9] Potential = A*exp(-Br) – Cr(-6) with unified parameters according to Table S5.

*Table S5. Unified (UNI) pair-potential parameters:*

| atom1 | atom2 | A | B | C |
|---|---|---|---|---|
| F7 | F7 | 170916.4 | 4.22 | 564.8 |
| F7 | O4 | 182706.1 | 3.98 | 868.3 |
| F7 | O5 | 182706.1 | 3.98 | 868.3 |
| F7 | C31 | 196600.9 | 3.84 | 1168.8 |
| F7 | H10 | 64257.8 | 4.11 | 248.4 |
| O4 | O4 | 195309.1 | 3.74 | 1335.0 |
| O4 | O5 | 195309.1 | 3.74 | 1335.0 |
| O4 | C31 | 393086.8 | 3.74 | 2682.0 |
| O4 | H10 | 295432.3 | 4.82 | 439.3 |
| O5 | O5 | 195309.1 | 3.74 | 1335.0 |
| O5 | C31 | 393086.8 | 3.74 | 2682.0 |
| O5 | H10 | 295432.3 | 4.82 | 439.3 |
| C31 | C31 | 226145.2 | 3.47 | 2418.0 |
| C31 | H10 | 120792.1 | 4.10 | 472.8 |
| H10 | H10 | 24158.0 | 4.01 | 109.2 |

## Supplementary figures

*Figure S7:* DSC traces for the new material studied: a) first heating cycle, to avoid decomposition the temperature range was limited to 190°C b) cooling cycle, c) second heating cycle.

*Figure S8*: Optical texture changes under applied dc electric field in SmA$_F$ (upper row) and SmC$_F$ (bottom rows) phases. In both phases, above a certain threshold voltage a homeotropic state is obtained, with polarization oriented along the applied field.

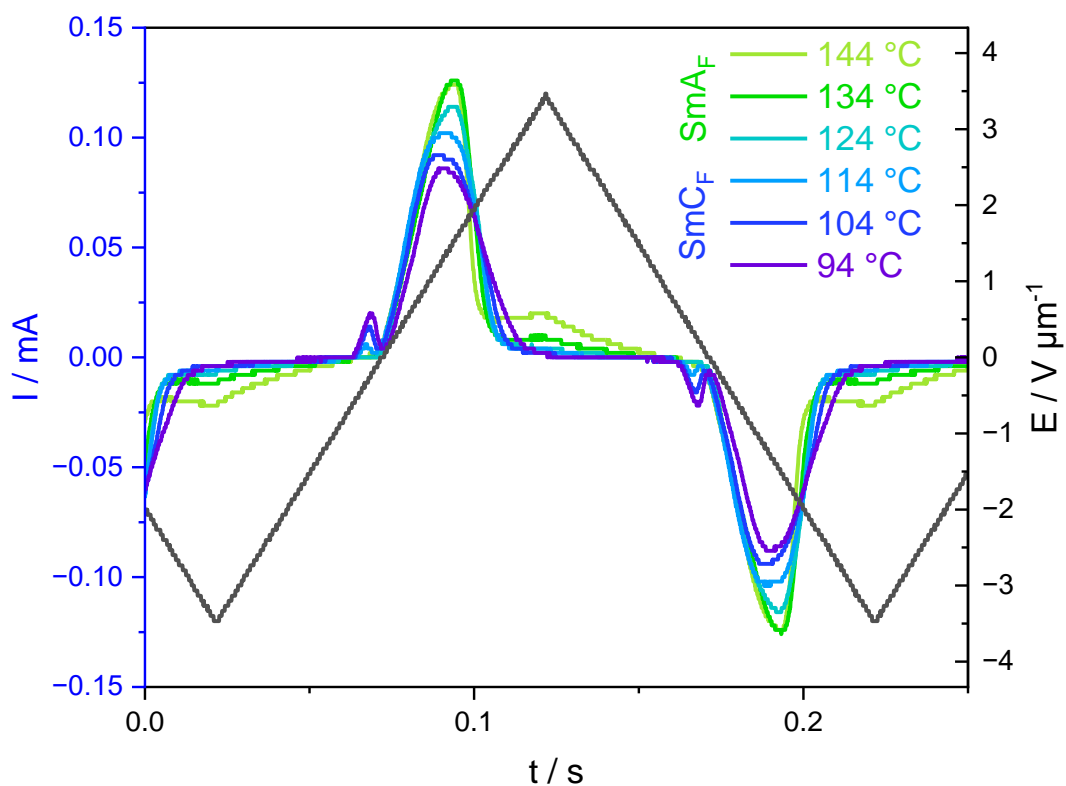

*Figure S9*: The switching current recorded in the SmC$_F$ and SmA$_F$ phases (blue and green lines) under application of triangular wave voltage (black line). On heating, the small peak observed at low voltages in the SmC$_F$ phase decreases and disappears on entering the SmA$_F$ phase. The main peak narrows and grows in height on heating through both smectic phases.

## Supplemental References